\documentclass[12pt,preprint]{aastex}

\def\L54{L_{54}}
\def\L52{L_{52}}
\def\et3{\eta_3}
\def\th1{\theta_{-1}}
\def\r7{r_{0,7}}
\def\x05{x_{0.5}}
\def\mathnew{\mathsurround=0pt}
\def\simov#1#2{\lower .5pt\vbox{\baselineskip0pt \lineskip-.5pt
       \ialign{$\mathnew#1\hfil##\hfil$\crcr#2\crcr\sim\crcr}}}
\def\simg{\mathrel{\mathpalette\simov >}}
\def\siml{\mathrel{\mathpalette\simov <}}
\def\Mesz{M\'esz\'aros~}
\def\bitm{\bibitem}

\def\L54{L_{54}}
\def\r07{r_{0,7}}
\def\et600{\eta_{600}}
\def\x05{x_{0.5}}
\def\rph{r_{ph}}
\def\beq{\begin{equation}}
\def\enq{\end{equation}}
\def\vareps{\varepsilon}
\def\eps{\epsilon}
\def\cm{\hbox{~cm}}
\def\s{\hbox{~s}}
\def\GeV{\hbox{~GeV}}
\def\MeV{\hbox{~MeV}}
\def\keV{\hbox{~keV}}
\def\eV{\hbox{~eV}}
\def\para{\parallel}
\def\bea{\begin{eqnarray}}
\def\ena{\end{eqnarray}}

\slugcomment{ApJL, subm. March 8, 2011; in press}

\shorttitle{Magnetic collisional GRBs}

\begin{document}

\title{GeV Emission from Collisional Magnetized Gamma Ray Bursts}
\author{P. \Mesz\footnote{
Dept. of Astronomy \& Astrophysics, Dept. of Physics and Center for Particle Astrophysics,
525 Davey Lab., Pennsylvania State University, University Park, PA 16802, USA} 
~~and 
M.J. Rees\footnote{Institute of Astronomy, University of Cambridge, Cambridge CB3 0HA, U.K.}}

\begin{abstract}
Magnetic fields may play a dominant role in gamma-ray bursts, and recent 
observations by {\it Fermi} indicate that GeV radiation, when detected, 
arrives delayed by seconds from the onset of the MeV component. Motivated by this,
we discuss a magnetically dominated jet model where both magnetic dissipation 
and nuclear collisions are important. We show that, for parameters typical of
the observed bursts, such a model involving a realistic jet structure can 
reproduce the general features of the MeV and a separate GeV radiation 
component, including the time delay between the two. The model also
predicts a multi-GeV neutrino component.

\end{abstract}



\section{Introduction}
\label{sec:intro}

Recent observations with the {\it Fermi} satellite have shown the presence of
GeV emission in GRB bursts, delayed by seconds relative to the usual MeV
radiation, e.g. \cite{abdo+09a,ackermann+10}. In some bursts, this delayed GeV
radiation appears as a distinct spectral component \citep{ackermann+10,abdo+09b}.
There have also been indications that the outflow in such bursts may be magnetized,
e.g. \cite{zhang+09mag,fan09}. Also on the theoretical side, there has been an 
increasing interest in magnetically dominated but baryon loaded GRB models, 
e.g. \cite{thompson94,drenkhahn+02,lyutikov+03,metzger+11,mckinney+11}. 

The fast rotating central engine, whether a black hole or a magnetar, 
can lead to a highly magnetized outflow which is initially Poynting dominated, 
with a sub-dominant baryon load at the base of the flow $r_0$ such that 
$\eta\equiv \left({\cal{E}}_{EM}/\rho_b c^2\right)_0 \gg 1$, where 
$\cal{E}_{EM}$ is electromagnetic (mainly magnetic) energy density, and
$\rho_b$ is baryon mass density. If the magnetic field in the jet is striped,
i.e. it has alternating polarity, either as a result of flow instabilities in 
a collapsar outflow or in oblique magnetar outflows, reconnection and dissipation 
can occur both below and above the photosphere, as discussed by \cite{giannios+07,
metzger+11,mckinney+11}, which can give rise to efficient radiation in the MeV band. 
Dissipative photospheres have also been proposed in magnetically sub-dominant baryon 
loaded outflows, as the source of the MeV radiation,  e.g. \cite{rees+05,peer+06,
ryde+10,peer+11,beloborodov11}, while the role of nuclear collisions in producing 
dissipation, high energy photons and neutrinos has been discussed by, e.g. 
\cite{meszaros+00nu,rossi+06,ioka10,beloborodov10} and others.

In this paper, we analyze a magnetically dominated and baryon loaded outflow,
and discuss the combined effects of dissipation from magnetic reconnection
and nuclear collisions in the context of the magnetically dominated outflow
dynamics, which differs from the usual hydrodynamic regime dynamics. We show
that such outflows can produce the usual MeV spectrum, as well as a GeV component
which arrives with a delay of seconds after the MeV component, for a minimally
(realistic) structured jet configuration. 

\section{Magnetic Dynamics, Nuclear Collisions and Photosphere}
\label{sec:dyn}

\noindent
{\it Dynamics}.-  
If the magnetic field is striped and the jet is one-dimensional\footnote{ 
For more general magnetic jet geometries and acceleration laws see text
below eq.(\ref{eq:rads}) and \cite{komissarov+09,tchek+10,mckinney+11}.}
the outflow accelerates \citep{drenkhahn02,granot+11,metzger+11} roughly as
\beq
\Gamma(r) \simeq \cases{(r/r_0)^{1/3} & for $r\leq r_{sat}$\cr
                                  \eta         & for $r\geq r_{sat}$\cr}
\label{eq:magac}
\enq
where $r_0=10^7\r07$ is the base of the outflow and $r_{sat}\simeq r_0 \eta^3$ 
is the radius at which the bulk Lorentz factor $\Gamma$ has saturated, the 
outflow having become approximately matter dominated. 
One way to roughly understand this dynamic behavior is as follows.
The comoving reconnection time is $t'_r \sim \lambda'/v'_r = \Gamma\lambda/v'_r
\sim \Gamma \eps_1 r_0/\eps_2 v_A \propto \Gamma$, where
$\lambda=\eps_1 r_0$ is the approximate radial lengthscale of field polarity
changes, $\eps_1\sim 1$, $v'_r \sim \eps_2 v'_A$ is reconnection velocity, 
$v'_A \sim c$ is the Alfv\'en velocity, with, e.g., $\eps_2\siml 0.1$ 
and comoving (lab) frame quantities are denoted as primed (unprimed).
The dynamic expansion time is $t'_{ex}\sim r/c\Gamma$. The comoving internal 
energy per baryon of the plasma (which is mainly magnetic, in the acceleration
phase) can be expressed as $\gamma'$. The outflow will be self-similar, and
$\gamma' \geq 1$, a dimensionless quantity, can be expressed as a dimensionless 
combination of dimensional quantities in the flow.  There are only two physically 
independent dynamical quantities, $t'_r$ and $t'_{ex}$,  suggesting $\gamma'\propto
t'_r /t'_{ex} \propto \Gamma^2/r$. The conversion of internal energy during the 
expansion is what leads to the growth of the bulk kinetic energy, and energy 
conservation requires $\gamma' \Gamma\sim {\rm const.}\propto \Gamma^3/r$,
or $\Gamma\propto r^{1/3}$. The final bulk Lorentz factor is the same $\Gamma\simeq 
\eta$ as in the pure hydrodynamic regime, but the growth is slower, because at each 
e-folding in radius only part of the magnetic internal energy gets converted directly 
into kinetic energy, the other part gets converted first into thermal energy, and 
only after another e-folding into kinetic energy. 

\noindent
{\it Longitudinal pion optical depth}.-
In a simple flow which initially moves with a single bulk Lorentz factor,
nuclear inelastic collisions couple $n$ and $p$ with a cross section 
$\sigma_{nuc}\sim \sigma_\pi (c/v_{rel})$, where $v_{rel}$ is the
relative drift velocity between $n$ and $p$ and $\sigma_\pi\sim 3\times
10^{-26}\cm^2$. When $v_{rel}\to c$, the nuclear collisions become inelastic
resulting in the production of pions, muons, and eventually $e^\pm$ pairs and
and neutrinos. The latter escape, but $\sim 1/4$ of the collision energy
ends up in pairs and photons, which are trapped in the optically thick flow,
adding to its thermal energy. 

In such simple homogeneous flows,  the nuclear collisions become inelastic 
(resulting in pions) when the comoving expansion time $t'_{ex}\sim r/c\Gamma$
becomes shorter than the comoving collision time $t'_{nuc}\sim 1/(n'_p \sigma_\pi c)$, 
or $\tau'_{\pi,\para} \sim {n'}_p \sigma_{nuc} r/\Gamma \sim 1$. 
Here $n'_p$ is the comoving proton density
\beq
n'_p = \frac{L x}{4\pi r^2 m_p c^3 \eta \Gamma},
\enq
where $L$ is the total kinetic luminosity, and $x=n_p/n_b \equiv n_p/(n_p+n_n)$
is the proton fraction of the baryon density $n_b=n_p+n_n$. The radius at which
pion production occurs is
\beq
\frac{r_\pi}{r_0} = \frac{L\sigma_\pi x}{4\pi m_pc^3 r_0}\frac{1}{\eta\Gamma^2}
     \equiv \eta_\pi^6 \frac{x}{\eta\Gamma^2},
\label{eq:rpigen}
\enq
where 
\beq
\eta_\pi \equiv \left(\frac{L\sigma_\pi}{4\pi m_p c^3 r_0}\right)^{1/6}
    \simeq 1.33\times 10^2 \L54^{1/6}\r07^{-1/6},
\label{eq:etapi}
\enq
and we assume parameters comparable to those deduced for typical bright LAT bursts,
e.g. GRB 080916C \citep{abdo+09a}, $L \sim 10^{54}\L54$, $\eta\simg 600\et600$, 
$\theta\sim 10^{-2}$.

The nuclear collisions can become inelastic in the regime $\Gamma \simeq 
(r/r_0)^{1/3} < \eta$, in which case from eqs. (\ref{eq:magac}, \ref{eq:rpigen}) 
the longitudinal pionization radius  is
\beq
\frac{r_\pi}{r_0}= x^{3/5} \eta_\pi^3 \left(\frac{\eta_\pi}{\eta}\right)^{3/5}
  ~~~{\rm for}~~~r_\pi<r_{sat},~~{\rm i.e.}~~\eta>\eta_\pi x^{1/6}=0.9\x05^{1/6} \eta_\pi.
\label{eq:rpimag}
\enq
The pionization radius $r_\pi$ and $\eta_\pi$  of eqs. (\ref{eq:rpimag},\ref{eq:etapi})
scale with $L,~r_0$ differently than in the corresponding hydrodynamic outflows, e.g. 
\cite{bahcall+00} and others, due to using here the magnetized outflow dynamics 
(\ref{eq:magac}) instead of $\Gamma \propto r$.  Beyond $r_\pi$, the neutrons decouple 
from the protons and start to coast with a terminal bulk Lorentz factor
\beq
\Gamma_n \simeq  x^{1/5} \eta_\pi(\eta_\pi/\eta)^{1/5}
        \simeq 80 ~\L54^{1/5}\et600^{-1/5}\x05^{1/5},
\enq
while the protons continue to accelerate as $\Gamma\propto r^{1/3}$ until reaching
the coasting value $\Gamma\sim \eta$. 
If on the other hand the flow Lorentz factor saturates before the protons and 
neutrons decouple, i.e. before the nuclear collisions become inelastic, which 
occurs if $\eta<\eta_\pi x^{1/6}$, the protons and neutrons continue to coast 
together at $\Gamma\sim \eta=$ constant and pionization never occurs.

\noindent
{\it Photon optical depth}.-
If the only electrons in the outflow are those associated with the protons, the
Thomson photosphere occurs at
\beq
\frac{\rph}{r_0} = \left(\frac{L \sigma_T}{4\pi m_p c^3 r_0}\right)\frac{1}{\eta\Gamma^2}
      = \eta_T^6 \frac{1}{\eta\Gamma^2}
\label{eq:rphgen}
\enq
where
\beq
\eta_T =  \left(\frac{L \sigma_T}{4\pi m_p c^3 r_0}\right)^{1/6}
      \equiv \eta_\pi \left(\frac{\sigma_T}{\sigma_\pi}\right)^{1/6}\simeq (5/3)\eta_\pi
      \simeq 2.22\times 10^2 ~\L54^{1/6}\r07^{-1/6}.
\label{eq:etast}
\enq
For $\eta >\eta_T ~(\eta <\eta_T$) the photosphere occurs at 
$r<r_{sat}~(r>r_{sat})$, at a radius
\beq
\frac{\rph}{r_0}=\cases{ 
     \eta_T^3 (\eta_T/\eta)^{3/5} & for~~$r<r_{sat}~~(\eta>\eta_T$);\cr
     \eta_T^3 (\eta_T/\eta)^3     & for~~$r>r_{sat}~~(\eta<\eta_T$).}
\label{eq:rph}
\enq
The scaling, as well as the definition of $\eta_T$ again differs from those
in e.g. \cite{meszaros+00ph} or \cite{beloborodov10}, due to our using here the 
magnetized outflow dynamics.
Thus, the photosphere is typically larger then the longitudinal pionosphere by a factor
\beq
\frac{\rph}{r_\pi}\sim \left(\frac{\sigma_T}{\sigma_\pi}\right)^{3/5}\sim 6.4.
\enq

The inelastic nuclear collisions produce additional leptons which can  contribute
to the Thompson opacity.  For a strongly magnetized outflow the initial secondary
leptons would cool very fast and the cascade efficiency drops as the pair 
production approaches its threshold, where in the limit \citep{beloborodov10,vurm+11} 
each $n,p$ collision adds at most one further lepton. Since $\tau_\pi \propto 
r^{-5/3}$ drops rapidly, most neutrons will suffer only one inelastic collision, 
increasing the electron density by a factor $\sim 2$. In addition, the reconnection 
process can also produce pairs, whose effects could be significant 
\citep{mckinney+11,metzger+11}. Assuming that the two effects contribute a 
comparable change in the lepton density, this increases $L$ by a factor $\gtrsim 4$ 
in eqs.(\ref{eq:rphgen},\ref{eq:etast}, \ref{eq:rph}). Thus, the photospheric radius 
could be a factor $\gtrsim 4^{3/5}\sim 2.3$ larger than its value without pairs in 
eqs.  (\ref{eq:rphgen},\ref{eq:rph}), i.e. $\rph \gtrsim 10-15 r_\pi$.

\noindent
{\it Transverse pion optical depth}.-
Neutrons, if present outside the jet, can drift sideways into the jet channel 
unhindered by magnetic fields. The lab-frame transverse nuclear collision optical 
depth of a jet of opening half-angle $\theta$ at radius $r$ is
\beq
\tau_{\pi,\perp}\sim n_p \sigma_\pi r\theta=
  \left(\frac{L\sigma_\pi}{4\pi m_pc^3 r_0}\right)\frac{r_0}{r}\frac{\theta}{\eta}
\enq
and the jet becomes transversely optically  thin at
\beq
\frac{r_{\pi,\perp}}{r_0}=\eta_\pi^6 \frac{\theta}{\eta}.
\enq
Thus, for a jet opening half-angle $\theta=10^{-2}\theta_{-2}$, we have the 
following radii,
\bea
r_\pi  \simeq & 6\times 10^{12} \L54^{3/5}\r07^{-3/5} \x05^{3/5} \et600^{-3/5} \cm,\cr
\rph   \simeq & 6\times 10^{13} \L54^{3/5}\r07^{-3/5} \et600^{-3/5} \cm, ~~~\cr
r_{\pi,\perp} \simeq & 9\times 10^{14} \L54 \et600^{-1} \theta_{-2} \cm,~~~~~~~~\cr
r_{sat} \simeq & 2\times 10^{15}\r07\et600^3 \cm~~~~~~~~~~~~~.
\label{eq:rads}
\ena

\noindent
{\it Transverse jet structure and neutron drift}.-
In a realistic jet the properties vary also in the transverse direction.
Hydrodynamical simulations (e.g. \citep{morsony+10} indicate a Lorentz factor
tapering off towards the edges, while MHD force-free simulations \citep{tchek+08}
show models where, depending on the stellar pressure profile and the magnetic 
symmetries, the jet opening angle can initially narrow with increasing radius, with
an inner jet core accelerating initially more slowly than the outer jet portions, 
but before exiting the star the inner jet becomes faster than the outer sheath, 
after which the jet opening angle remains constant. 
As a simple idealized model, we can consider the transverse
structure outside the star as a two-step jet, consisting of an inner jet core with,
e.g., a nominal constant half-angle $\theta=10^{-2}\theta_{-2}$ with $\eta=600\et600$, 
and a slower\footnote{Depending on various parameters an alternative model might 
have outside the star a slower inner jet and a faster sheath, which could lead to 
results similar to those discussed below; we do not consider this case here.}
outer jet or sheath with, say, $\eta_{out}\sim 100$, where both 
inner core and outer sheath have been populated with protons and free neutrons
already near the black hole, where most nuclei get photo-dissociated 
\citep{beloborodov03,metzger+08}.
One can expect neutrons from the outer sheath to drift into  the jet core, and 
for $\eta_{out}\siml 10^2\sim \theta^{-1}$ the neutrons 
can penetrate transversely the entire jet core. Also, the relative radial Lorentz 
factor ratio between indrift neutrons and jet core baryons (either $p$ or $n$) is 
$\Gamma_{rel}\sim \eta/\eta_{out} > 1$, so the collisions will be inelastic, leading 
again to pions.

Over a timescale $t$ the thermal neutrons from an outer sheath where the lab-frame 
neutron density is $n_n$ and the thermal neutron velocity is $v_t$ will diffuse 
transversely into the jet, their flux being 
\beq
\phi_n \sim \left(\frac{v_t \ell}{t}\right)^{1/2} n_n 
          \sim  \beta_t  \left(\frac{n_n c}{\sigma_\pi t}\right)^{1/2},
\label{eq:phin}
\enq
where $\ell\sim 1/(n_n\sigma_{nuc})$, $\sigma_{nuc}\sim \sigma_\pi(c/v_t)$ and
$\beta_t =v_t /c$. Taking a jet transverse area of $A\sim \pi\theta r^2$, on a
timescale $t$ the total number of neutrons drifting transversely into the jet core 
will be, roughly,
\beq
N_{n,\perp}\sim \pi \theta r^2 \phi_n t 
      \sim \pi\theta r^2 ~\beta_t (n_n ct/\sigma_\pi)^{1/2},
\label{eq:Nn}
\enq
while over the same timescale the number of baryons ($n$ and $p$) passing 
longitudinally through the jet core is
\beq
N_{b,\para}\sim \pi \theta r^2 n_b ct 
    \sim \pi\theta r^2 ~(L/4\pi r^2 m_pc^3\eta) ct
\label{eq:Np}
\enq
The number of collisions is maximized  when $N_{n,\perp} \gtrsim N_{b,\para}$,
that is $ \beta_t \left(n_n {ct}/{\sigma_\pi}\right)^{1/2} \gtrsim 
\left({ct}/{\sigma_\pi}\right) \left({\eta_\pi^6 r_0}/{r^2\eta}\right)$,
and when the jet transverse pion optical depth $\tau_{\pi,\perp}\sim 1$, i.e. 
$r\sim r_{\pi,\perp}$.  For the first condition, we can estimate the outer sheath
neutron thermal velocity $\beta_t$ since we know that the comoving baryon 
temperature becomes trans-relativistic ($T\sim$ GeV) at the saturation radius
$r_{sat,out}=\r07\eta_{out}^3=10^{13}\cm$, where $v_t\sim c$, after which they 
coast with with $T_t\propto r^{-2/3}$, so at $r_{\pi,\perp}$ we have $\beta_t^2
=(v_t/c)^2 \sim (kT_t/m_pc^2)= (r_{sat,out}/r_{\pi,\perp})^{-2/3} \sim 10^{-4/3}$
These two conditions are then met when at $r\sim r_{\pi,\perp}$ the 
neutron density in the outer sheath is
\beq
n_n \gtrsim 
\frac{ct}{\sigma_\pi}\frac{\eta^2}{\eta_\pi^{12}}\frac{1}{r_0^2 \theta^4\beta_t^2}
 \sim 4\times 10^{11} \r07^{-2}\theta_{-2}^4\L54^{-2}\et600^2 t \cm^{-3}
\label{eq:nn}
\enq
over a timescale $t$. The density in eq. (\ref{eq:nn}) is comparable to what might 
be expected in an outer jet sheath. 

\section{Spectrum Formation}
\label{sec:sp}

\noindent
{\it GeV radiation}.-
The `prompt' component of the GeV radiation is expected to arise from the
transverse drift nuclear collision mechanism at $r_{\pi,\perp}$ discussed above.
The $\gamma\gamma$ photosphere to $E=10\GeV$ photons can be roughly estimated as 
\citep{beloborodov10} $r_{\gamma\gamma}(E)\sim 10^{15}(E/10\GeV)^{3/2} 
\et600^{-5}\L54\cm$ , so that
\beq
\rph < r_{\gamma\gamma}(10\GeV) \siml r_{\pi,\perp},
\label{eq:rcomp}
\enq
the right inequality holding for $\et600\simg 1$.
Thus, at $r_{\pi,\perp}$, which is at or outside $r_{\gamma\gamma}$ for $\et600\simg 1$,
multi-GeV photons will be copiously produced by transverse indrift neutron collisions 
with jet core baryons, and these photons can escape unhindered.
The energy in this GeV component, for the conditions discussed in the previous section,
is a significant fraction of the kinetic luminosity.
Most pions will be formed approximately at rest in the frame of the jet, 
and the processes $p,N \to \pi^+,\pi^-,\pi^0$, where $N=(p,n)$, occur in 
approximately equal ratios. Thus 1/3 of the collisions on average yield two photons 
from $\pi^0\to 2\gamma$ decays at an observer frame energy which is broadly 
distributed around a central energy 
$\vareps^{\pi,\perp}_\gamma\sim \eta (m_\pi c^2/2)/(1+z) \sim 7~\et600 (5/1+z)\GeV$, 
extending up to twice that value and down to a fraction of it.

\noindent
{\it GeV time delay}.-
The relative time delay between between MeV photons from the central photosphere
(see below) and the observed onset of the GeV emission, which arises at $r_{\pi,\perp}$ 
from neutrons from an outer jet sheath moving with bulk Lorentz factor $\eta_{out}$ 
which drift into the inner jet at the radius $r_{\pi,\perp}$ where their effect becomes
significant is
\beq
\Delta t\sim \frac{r_{\pi,\perp}}{2 c\eta_{out}^2}
      \sim  1~\L54 \r07\et600^{-1}\eta_{out,100}^{-2}~\s,
\label{eq:tdel}
\enq
in the lab frame at the source. In the observer frame this is lengthened by a factor
$(1+z)$. For a burst at $z\sim 4$ such as GRB 080916C this is similar to the observed 
GeV-MeV delay of $\sim 5$ seconds observed by Fermi in that object.

\noindent
{\it MeV radiation}.-
The MeV (Band) spectrum may be expected to arise from the dissipative photosphere 
of the inner jet at $\rph \sim 6\times 10^{13} \cm$ (eq.[\ref{eq:rads}]). 
This MeV spectrum may be attributed to synchrotron radiation following magnetic or 
shock dissipation near $\rph$, with typical bulk Lorentz factors $\Gamma_r\sim 1$. 
This would result in random magnetic fields $B'\sim (32\pi \eps_B m_p c^2 n_b')^{1/2}
\Gamma_r$, where $n'_b=L/(4\pi \rph^2 \eta\Gamma_{ph})$ is the baryon comoving
density at $\rph$, where for the nominal jet parameters used we have $\Gamma_{ph}=
(\rph/r_0)^{1/3}\sim 180\L54^{1/5}\r07^{-8/15}\et600^{-1/5}$, so $n'_b\sim 
5\times 10^{12}\L54^{-2/5}\r07^{26/15}\et600^{2/5}$ and $B'\sim 4\times 10^5
\L54^{-1/5}\r07^{13/15}\et600^{1/5}\eps_B^{1/2}\Gamma_r$ G. The same magnetic 
dissipation leading to semi-relativistic shocks with comoving bulk Lorentz factors
$\Gamma_r\sim 1$ near  $\rph$ will accelerate electrons to average non-thermal random 
Lorentz factors of $\gamma'_e\sim (m_p/m_e)\Gamma_r \sim 10^3\Gamma_r$, assuming 
equipartition between baryons and electrons, leading to an observer frame synchrotron 
radiation peak at $\vareps^{ph}_{sy} =\vareps'_{sy}\Gamma_{ph}(1+z)^{-1}
\sim 0.3~\r07^{1/3}\eps_B^{1/2}\Gamma_r^3 (5/1+z)\MeV$. This non-thermal component 
will dominate over a much weaker photospheric thermal component at $kT_{th}\sim 1\keV$.
Comptonization may boost a fraction of the synchrotron photons to higher energies, 
but using the expression above for $r_{\gamma\gamma}$ a cutoff is obtained at
$\vareps_{\gamma\gamma}\sim 3~\MeV \L54^{-2/3} \Gamma_{ph,180}^{10/3}$.  Thus, the 
photosphere allows escape of MeV but not GeV photons. A decreasing fraction of 
increasing energy photons could arise from dissipation at larger radii 
\citep{beloborodov10}, as $r_{\gamma\gamma}$ increases, but significant GeV escape
compatible with a second spectral component is only expected from radii where 
transverse neutron drift is important.

\noindent
{\it Outer jet photosphere}.-
The outer jet has a photosphere beyond its saturation radius, at $r_{ph,out}\sim 
2.5\times 10^{14}\L54\eta_{out,100}^{-3}\cm$. If there was dissipation at this
outer photosphere, a calculation similar to the inner jet's would give a
synchrotron peak at $\vareps_{sy,out}\sim 25~\L54^{-1/2}\eps_B^{1/2}\Gamma_r^3
\eta_{out,100}^{1/2}(5/1+z)\keV$. 
However, magnetic dissipation is not expected to occur above the saturation radius,
and neither does nuclear collisional dissipation, since the $n,p$ components do not 
decouple for this $\eta<\eta_\pi\sim 130$. Thus, unless some other form of dissipation
occurs near the outer photosphere, it would have a quasi-blackbody spectrum with 
$kT\sim 0.2~\L54^{-1/2}\r07^{1/6}\eta_{out,100}^{2/3}(5/1+z)~\keV$ and a luminosity 
$L_{ph,out}\sim \eps_{rad}L_0(r_{sat}/\rph)^{2/3} \siml 10^{-2} (\eps_{rad}/0.5)L_0$, 
well below the kinetic luminosity $L_0$.

\noindent
{\it Internal shocks}.- In principle these would not be expected if the magnetization 
remains significant beyond the saturation radius (c.f. \cite{zhang+11}). If they did
occur, the earliest would need to be at $r_{IS}\sim 2ct_v\eta^2 \simg 2\times 10^{15}
t_{v,-1} \et600^2\cm \simg r_{sat}$, where $t_v=10^{-1}t_{v,-1}\s$ is the variability 
time, and the usual arguments would lead to a synchrotron peak at $\vareps^{IS}_{sy}
\siml 8~\L54^{1/2}\eps_B^{1/2}\eps_e^2 t_{v,-1}^{-1}\et600^{-1}\Gamma_r(5/1+z)\keV$.

\noindent
{\it External shocks and afterglow}.-
Other radiation components are due to the external shocks of the inner and outer 
jets. Forward shocks (FS) are guaranteed, while reverse shocks (RS) are less certain,
depending on the residual magnetization $\sigma$ value at the external shock.
For an outflow (prompt burst) duration $t=10t_1\s$ in the burst frame, in an
external medium of density $n_0\cm^{-3}$, the inner jet external shock occurs at a 
deceleration radius 
$r_d\sim 1.6\times 10^{17} \L54^{1/3} t_1^{1/3}n_0^{-1/3}\et600^{-2/3}\cm$
at an observer time 
$t_d\sim 50 \L54^{1/3}t_1^{1/3}n_0^{-1/3}\et600^{-8/3}([1+z]/5)\s$.
With the usual assumption about magnetic field and electron energy equipartition, and
$\Gamma_d\sim \eta/2$ at $r_d$, the FS synchrotron  spectrum of the inner jet peaks at 
$\vareps^{FS}_{sy}\sim 2\eps_B^{1/2}\eps_e^2\et600^4 (5/1+z)\MeV$. For the usual
power law electron distribution accelerated at the shock, this synchrotron spectrum 
extends as a power law into the GeV range \citep{ghiselliini+10,kumar+10}, providing 
the long term GeV afterglow, the external shock being well outside 
$r_{\gamma\gamma}(10\GeV)$. The earliest prompt GeV radiation, however, appears
to require a separate origin, as considered by \cite{toma+11,he+11,depasquale+10,
corsi+10} in the context of non-magnetic outflows.

The outer jet deceleration radius is 
$r_{d,out}\sim 5.3\times 10^{17} \L54^{1/3} t_1^{1/3}n_0^{-1/3}\eta_{100,out}^{2/3}\cm$ 
at an observer time 
$t_{d,out}\sim 2\times 10^3~\L54^{1/3}t_1^{1/3}n_0^{-1/3}\eta_{100,out}^{-8/3}([1+z]/5)\s$,
and the outer jet FS synchrotron spectral peak is at 
$\vareps^{FS}_{sy,out}\sim 2~\eps_B^{1/2}\eps_e^2\eta_{out,100}^4(5/1+z)~\keV$.

If the magnetization parameter of the ejecta $\sigma$ has become small at $r_d$,
a reverse shock may form \citep{mimica+09,zhang+11}. With the usual assumptions 
for an equipartition magnetic field in a reverse shock with $\Gamma_r \sim 1$ and 
$\gamma'_e\sim \eps_e 10^3\Gamma_r$ the inner jet RS synchrotron peak occurs at 
$\vareps^{RS}_{sy} \sim 70~\L54^{1/6}t_1^{1/3}n_0^{1/2}\eps_B^{1/2}\eps_e^2
\Gamma_r^3\et600^{2/3}(5/1+z)~\eV$, while the outer jet reverse shock synchrotron 
spectrum peaks at $\vareps^{RS}_{sy,out}\sim 20~\eV$, with the same scaling.

\noindent
{\it Neutrino spectrum}.-
Also, as a result of the charged pion decays, 1/3 of the collisions yield on average 
a muon neutrino of energy broadly centered around $\vareps_\nu\sim 7~ \et600 (5/1+z)\GeV$ 
(again extending up to twice that value and down to a fraction of it), and a muon neutrino 
as well as an electron neutrino centered at energy $\vareps_\nu\sim 3~\et600 (5/1+z)\GeV$, 
while another 1/3 of the collisions leads to the corresponding antineutrinos of similar 
energies.  These are in the sensitivity range of the Deep Core array of the IceCube 
neutrino detector.

\section{Discussion}
\label{sec:disc}

We have presented a GRB model based on a magnetically dominated, baryon loaded
outflow, where both magnetic reconnection and nuclear $p,n$ collisions lead
to dissipation around and above a photon photosphere. We have discussed the
dynamics of the magnetized jets based on a simplified reconnection prescription 
and an idealized jet transverse structure, deriving the radii at which the
photon scattering photosphere occurs as well as the radii at which radial 
and transverse drifts between proton and neutron components lead to nuclear
collisional dissipation, in the context of the magnetic jet dynamics. 
Although the detailed structure and dynamics could depend on various
assumptions about the stellar envelope pressure profile, the magnetic  field
configuration, etc. \citep{tchek+08,mckinney+11,metzger+11}, the present model 
qualitatively captures a number of the essential features. In particular, 
the characteristic radii and their scalings differ from those in a purely 
hydrodynamical jet. Both in this and in more general magnetic models a slower 
acceleration rate generically leads to larger photospheric and dissipation radii 
relative to the hydrodynamic case, in line with previous hints from {\it Fermi} 
observational results indicating larger emission radii.  

Using parameters inferred from {\it Fermi} observations of bursts showing significant 
emission in the LAT instrument in the GeV energy range,  which typically are of very 
high luminosity and higher than average Lorentz factor \citep{zhang+11b}, we discussed 
the gross properties of the MeV and GeV photon production spectra. For a simplified 
structured jet model consisting of a central faster jet core and a slower outer jet 
surrounding it, we showed that for reasonable parameters a significant GeV component 
is produced, with a time delay of seconds relative to the onset of the MeV component,
in rough agreement with the observations of GRB 080916C, e.g. \citep{abdo+09a}. 
Related to the photon emission, a multi-GeV neutrino emission component is also
expected, whose luminosity is comparable to that of the multi-GeV photon emission.
More detailed calculations, which are beyond the object of this letter, 
will be needed in order to make detailed spectral fits to specific bursts.

\acknowledgements{We thank NASA NNX09AL40G, NSF PHY-0757155, the Royal
Society, and the Institute of Astronomy, Cambridge, for partial support, and
Dr. K. Toma and the referee for useful comments.}

\end{document}